\documentclass[reprint,twocolumn,superscriptaddress,showkeys,
nofootinbib,notitlepage,amsmath,amssymb,floatfix]{revtex4-1}

\pdfoutput=1
\usepackage{latexsym,amsmath,amssymb,lmodern,float,url}
\usepackage{natbib}
\usepackage{color}
\usepackage{multirow}
\usepackage{bm}
\usepackage{array}

\usepackage{mathtools}

\def\de{\delta^{\vphantom{1}}}
\def\bde{{\bar\delta}}
\def\qq{{q\bar q^\prime}}
\def\QQ{{Q\bar Q}}
\def\kqQc{{\kappa_{qc}}}
\def\kqQs{{\kappa_{sc}}}
\def\kqQ{{\kappa_{qQ}}}
\def\ksQ{{\kappa_{sQ}}}

\def\QQqq{{Q\bar{Q}q\bar{q}}}

\def\ccss{{c\bar{c}s\bar{s}}}
\def\QQss{{Q\bar{Q}s\bar{s}}}
\def\ccqq{{c\bar{c}q\bar{q}^\prime}}

\def\QQqs{{Q\bar{Q}q\bar{s}}}

\def\bt{{\bar\theta}}

\def\h3{{\displaystyle{\frac 3 2}}}

\begin{document}
\title{Spectrum of Hidden-Charm, Open-Strange Exotics in the
Dynamical Diquark Model}
\author{Jesse F. Giron}
\email{jfgiron@asu.edu}
\author{Richard F. Lebed}
\email{Richard.Lebed@asu.edu}
\author{Steven R. Martinez}
\email{srmart16@asu.edu}
\affiliation{Department of Physics, Arizona State University, Tempe,
AZ 85287, USA}
\date{June, 2021}

\begin{abstract}
The observation by BESIII and LHCb of states with hidden charm and
open strangeness ($c\bar c q\bar s$) presents new opportunities for
the development of a global model of heavy-quark exotics.  Here we
extend the dynamical diquark model to encompass such states, using
the same values of Hamiltonian parameters previously obtained from
the nonstrange and hidden-strange sectors.  The large mass splitting
between $Z_{cs}(4000)$ and $Z_{cs}(4220)$ suggests substantial
SU(3)$_{\rm flavor}$ mixing between all $J^P \! = \! 1^+$ states,
while their average mass compared to that of other sectors offers a
direct probe of flavor octet-singlet mixing among exotics.  We also
explore the inclusion of $\eta$-like exchanges within the states, and
find their effects to be quite limited.  In addition, using the same
diquark-mass parameters, we find $P_c(4312)$ and $P_{cs}(4459)$ to
fit well as corresponding nonstrange and open-strange pentaquarks.
\end{abstract}

\keywords{Exotic hadrons, diquarks}
\maketitle

\section{Introduction}

The spectrum of known heavy-quark exotic hadrons continues to expand
frequently, with about 50 candidates observed to date.  Almost all
have a valence light-flavor content consisting of only
$q \! \equiv \! u$ and $d$ quarks, but very recently some states with
open strangeness have been detected in both the
open-charm~\cite{Aaij:2020hon,Aaij:2020ypa} ($\bar c d\bar s u$) and
hidden-charm sectors.  In the latter, both a
pentaquark~\cite{Aaij:2020gdg} ($c\bar c uds$) state and
tetraquark~\cite{Ablikim:2020hsk,Aaij:2021ivw} ($c\bar c u\bar s$)
states have been observed.\footnote{Charge-conjugate states are
implied as included throughout.}

Multiple reviews summarizing both experimental and theoretical
advances in this field have appeared in recent
years~\cite{Lebed:2016hpi,Chen:2016qju,Hosaka:2016pey,
Esposito:2016noz,Guo:2017jvc,Ali:2017jda,Olsen:2017bmm,
Karliner:2017qhf,Yuan:2018inv,Liu:2019zoy,Brambilla:2019esw}.
Several competing theoretical frameworks (di-hadron molecular states,
bound states of diquarks, threshold enhancements, {\it etc.}) have
been developed for years, but no single scheme has yet emerged as a
dominant paradigm to explain all the new states, analogous to the way
that quark-potential models successfully elucidate the conventional
$c\bar c$ and $b\bar b$ sectors~\cite{Zyla:2020zbs}.

Since a number of the observed exotics decay to final states like
$J/\psi \, \phi$ or $D_{s J}^{(*)} \bar D_{s J}^{(*)}$, they possess
a presumptive $c\bar c s\bar s$ valence content.  The advent of
$\bar c c\bar s q$ states thus introduces an intermediate case
between $c\bar c q\bar q^\prime$ and $c\bar c s\bar s$ cases, and
therefore not only provides an opportunity to examine whether a
particular theoretical picture can successfully incorporate data from
all of these flavor sectors, but also examines the manifestation of
light-quark SU(3)$_{\rm flavor}$ for the first time outside of
conventional mesons and baryons.

The new data in the hidden-charm, open-strange sector itself is quite
interesting.  BESIII observes a structure~\cite{Ablikim:2020hsk} in
the $K^+$ recoil spectrum of
$e^+ e^- \! \to \! K^+ (D_s^- D^{*0} \! + \! D_s^{*-} D^0)$ near the
2-charmed-meson thresholds (about 3975 and 3977~MeV, respectively).
For this $Z_{cs}(3985)$ state, they obtain:
\begin{eqnarray}
m_{Z_{cs}^-} & = & 3982.5^{+1.8}_{-2.6} \pm 2.1 \ {\rm MeV} \, ,
\nonumber \\
\Gamma_{Z_{cs}^-} & = & 12.8^{+5.3}_{-4.4} \pm 3.0 / {\rm MeV} \, ,
\end{eqnarray}
and $J^P \! = \! 1^+$.  Meanwhile, LHCb reports 2
states~\cite{Aaij:2021ivw} decaying to $J/\psi \, K^+$,
\begin{eqnarray} \label{eq:LHCbZcs}
m_{Z_{cs}^+ (4000)} & = & 4003 \pm 6^{+ \ 4}_{-14} \ {\rm MeV} \, ,
\nonumber \\
\Gamma_{Z_{cs}^+ (4000)} & = & 131 \pm 15 \pm 26 \ {\rm MeV} \, ,
\nonumber \\
m_{Z_{cs}^+ (4220)} & = & 4216 \pm 24^{+43}_{-30} \ {\rm MeV} \, ,
\nonumber \\
\Gamma_{Z_{cs}^+ (4220)} & = & 233 \pm 52^{+97}_{-73} \ {\rm MeV} \,
,
\end{eqnarray}
with $Z_{cs}(4000)$ carrying $J^P \! = \! 1^+$ and $Z_{cs}(4220)$
favored to carry $J^P \! = \! 1^+$.  The masses of $Z_{cs}(3985)$ and
$Z_{cs}(4000)$ are compatible with them being the same state, but
their measured widths are wildly different.  For the purposes of this
paper, we assume that only a single light $Z_{cs}$ state exists near
4.0~GeV, the discrepancy in width measurements perhaps arising from
effects caused by interactions with the nearby charmed-meson
thresholds.\footnote{This opinion is not universal.  For example,
Ref.~\cite{Maiani:2021tri} treats $Z_{cs}(3985)$ and $Z_{cs}(4000)$
as separate states belonging to distinct SU(3)$_{\rm flavor}$
multiplets.}  LHCb also observes a baryonic
structure~\cite{Aaij:2020gdg} $P_{cs}(4459)$ decaying to
$J/\psi \, \Lambda$:
\begin{eqnarray} \label{eq:Pcs}
m_{P_{cs}(4459)} & = & 4458.8 \pm 2.9^{+4.7}_{-1.1} \ {\rm MeV} \, ,
\nonumber \\
\Gamma_{P_{cs}(4459)} & = & 17.3 \pm 6.5^{+8.0}_{-5.7} \ {\rm MeV}
\, ,
\end{eqnarray}
although its $J^P$ value is not yet determined.

Quite an extensive body of theoretical work on the hidden-charm,
open-strange hadrons exists.  For example, in the meson sector whose
study forms the bulk of this work, several papers~\cite{Lee:2008uy,
Ebert:2008kb,Chen:2013wca,Dias:2013qga,Voloshin:2019ilw,
Ferretti:2020ewe} predate the experimental observations, while
multiple studies followed the announcement of the BESIII result but
preceded the appearance of the LHCb paper~\cite{Wang:2020kej,
Yang:2020nrt,Meng:2020ihj,Liu:2020nge,Wan:2020oxt,Chen:2020yvq,
Du:2020vwb,Cao:2020cfx,Sun:2020hjw,Wang:2020rcx,Wang:2020htx,
Wang:2020iqt,Azizi:2020zyq,Jin:2020yjn,Sungu:2020zvk,Ikeno:2021ptx,
Xu:2020evn,Meng:2020cbk,Wang:2020dgr,Guo:2020vmu,Albuquerque:2021tqd,
Ozdem:2021yvo,Yan:2021tcp}, and yet others appeared subsequent to the
LHCb results~\cite{Maiani:2021tri,Ge:2021sdq,Chen:2021uou,
Ortega:2021enc,Chen:2021erj,Maiani:2021dzz,Meng:2021rdg,
Ozdem:2021hka,Shi:2021jyr}.  As one may imagine, this body of work
encompasses multiple approaches, including molecular and diquark
models, chiral-quark models, and QCD sum rules, among others.

The present work uses the dynamical diquark model, initially
introduced in Ref.~\cite{Brodsky:2014xia} as a theoretical picture to
explain how relatively compact color-triplet diquark quasiparticle
pairs can form spatially extended tetraquark states, and extended in
Ref.~\cite{Lebed:2015tna} to describe pentaquarks as color-triplet
diquark-triquark quasiparticle bound states.  The picture is
developed in Ref.~\cite{Lebed:2017min} into a predictive model by
noting that the static interaction potential between the heavy
color-triplet quasiparticles is the same one as appearing in lattice
simulations of heavy quarkonium and its hybrids.  The multiplet band
structure for $c\bar c q\bar q^\prime$ and $c\bar c qqq$ states is
studied numerically in Ref.~\cite{Giron:2019bcs}; the fine structure
of the ground-state ($S$-wave) $c\bar c q\bar q^\prime$ multiplet is
examined numerically in Ref.~\cite{Giron:2019cfc} and that of the
$P$-wave multiplet appears in Ref.~\cite{Giron:2020fvd}.  An
analogous study of the $b\bar b q\bar q^\prime$ and $c\bar c s\bar s$
systems is presented in Ref.~\cite{Giron:2020qpb}, and the
$c\bar c c\bar c$ states are investigated in
Ref.~\cite{Giron:2020wpx}.  Radiative transitions between exotic
states are computed in Ref.~\cite{Gens:2021wyf}.

Our analysis of $c\bar c q\bar s$ exotics here directly interpolates
between the analysis of $c\bar c q\bar q^\prime$ in
Ref.~\cite{Giron:2019cfc} and $c\bar c s\bar s$ in
Ref.~\cite{Giron:2020qpb}, and uses the same numerical inputs.
However, we find that a careful treatment of the SU(3)$_{\rm flavor}$
structure introduces one new parameter, related to octet-singlet
mixing.  In addition, we allow for the possibility of an $\eta$-like
exchange between the diquarks analogous to the $\pi$-like exchange
already present in the original model~\cite{Giron:2019cfc}, but show
that its effects are quite limited by constraints from the
phenomenology of the $c\bar c q\bar q^\prime$ sector.  We find that
the known phenomenology of the open-strange sector does indeed follow
from that of the other sectors, despite superficially appearing quite
different.  We also carry out an analogous exercise for nonstrange
and open-strange hidden-charm pentaquark states [$P_{cs}(4459)$
currently being the only known example of the latter], and obtain
remarkably satisfactory results.

This paper is organized as follows.  In Sec.~\ref{sec:States} we
define the multiplets of states in terms of eigenstates of both good
diquark spin and good heavy-quark/light-quark spins.
Section~\ref{sec:Ham} presents the Hamiltonian for the
$c\bar c q\bar q^\prime$ and $c\bar c s\bar s$ sectors, now including
a possible term from $\eta$-like exchanges, and computes all relevant
matrix elements.  Section~\ref{sec:OpenStrange} performs the same
analysis for the $c\bar c q\bar s$ sector, and discusses possible
mixing between multiplets whose nonstrange members carry opposite $C$
parity.  In Sec.~\ref{sec:Analysis} we discuss the effects of
octet-singlet mixing on the analysis and present numerical results,
and in Sec.~\ref{sec:Concl} we summarize and conclude.

\section{States of the Model}
\label{sec:States}

A cataloguing of the $\QQ q\bar q^\prime$ or
$\QQ q \hspace{1pt} q_1 q_2$ states in the dynamical diquark model,
where $q, q^\prime \! , q_i \! \in \! \{ u, d \}$, first appears in
Ref.~\cite{Lebed:2017min}.  The same notation, with small
modifications, is applied to $c\bar c s\bar s$ in
Ref.~\cite{Giron:2020qpb} and to $c\bar c c\bar c$ in
Ref.~\cite{Giron:2020wpx}.  All confirmed exotic candidates to date
have successfully been accommodated within the lowest ($\Sigma^+_g$)
Born-Oppenheimer potential of the gluon field connecting the heavy
diquark [$\de \! \equiv (Qq)$]-antidiquark
[$\bde \! \equiv \! (\bar Q \bar q^\prime) $] or diquark-triquark
[$\bt \! \equiv \! (\bar Q (q_1 q_2))$] quasiparticles.  In all
cases, $\de, \bde, \bt$ are assumed to transform as color triplets
(or antitriplets) and each quasiparticle contains no internal orbital
angular momentum.

In the case of $\QQ q\bar q^\prime$, the classification scheme then
begins with 6 possible core states in which the quasiparticle pair
lie in a relative $S$ wave.  Indicating the total spin $s$ of a
diquark $\de, \bde$ by $s_\de, s_\bde$ and using a subscript on the
full state to indicate its total spin, one obtains the spectrum
\begin{eqnarray}
J^{PC} = 0^{++}: & \ & X_0 \equiv \left| 0_\de , 0_\bde \right>_0 \,
, \ \ X_0^\prime \equiv \left| 1_\de , 1_\bde \right>_0 \, ,
\nonumber \\
J^{PC} = 1^{++}: & \ & X_1 \equiv \frac{1}{\sqrt 2} \left( \left|
1_\de , 0_\bde \right>_1 \! + \left| 0_\de , 1_\bde \right>_1 \right)
\, ,
\nonumber \\
J^{PC} = 1^{+-}: & \ & \, Z \  \equiv \frac{1}{\sqrt 2} \left( \left|
1_\de , 0_\bde \right>_1 \! - \left| 0_\de , 1_\bde \right>_1 \right)
\, ,
\nonumber \\
& \ & \, Z^\prime \equiv \left| 1_\de , 1_\bde \right>_1 \, ,
\nonumber \\
J^{PC} = 2^{++}: & \ & X_2 \equiv \left| 1_\de , 1_\bde \right>_2 \,
.
\label{eq:Swavediquark}
\end{eqnarray}
Since 4 quark angular momenta are being combined, one may transform
these states into other convenient bases by means of $9j$ angular
momentum recoupling coefficients.  In particular, in the basis of
good total heavy-quark ($\QQ$) and light-quark ($\qq$) spin, the
transformation reads
\begin{eqnarray}
&&\left< (s_q \, s_{\bar q}) s_\qq , (s_Q \, s_{\bar Q}) s_\QQ
, S \, \right| \left. (s_q \, s_Q) s_\de , (s_{\bar q} \, s_{\bar Q})
s_\bde , S \right> \nonumber\\
&&=\left( [s_\qq] [s_\QQ] [s_\de] [s_\bde] \right)^{1/2}
\begin{Bmatrix}
s_q & s_{\bar q} & s_\qq \\
s_Q & s_{\bar Q} & s_\QQ \\ 
s_\de & s_\bde & S
\end{Bmatrix}
\, , \ \ \label{eq:9jTetra}
\end{eqnarray}
with $[s] \! \equiv \! 2s + 1$ signifying the multiplicity of a
spin-$s$ state.  Using Eqs.~(\ref{eq:Swavediquark}) and
(\ref{eq:9jTetra}), one then obtains
\begin{eqnarray}
J^{PC} = 0^{++}: & \ & X_0 = \frac{1}{2} \left| 0_\qq , 0_\QQ
\right>_0 + \frac{\sqrt{3}}{2} \left| 1_\qq , 1_\QQ \right>_0 \, ,
\nonumber \\
& & X_0^\prime = \frac{\sqrt{3}}{2} \left| 0_\qq , 0_\QQ
\right>_0 - \frac{1}{2} \left| 1_\qq , 1_\QQ \right>_0 \, , 
\nonumber \\
J^{PC} = 1^{++}: & \ & X_1 = \left| 1_\qq , 1_\QQ \right>_1 \, ,
\nonumber \\
J^{PC} = 1^{+-}: & \ & \, Z \; = \frac{1}{\sqrt 2} \left( \left| 
1_\qq , 0_\QQ \right>_1 \! - \left| 0_\qq , 1_\QQ \right>_1 \right)
\, , \nonumber \\
& \ & \, Z^\prime = \frac{1}{\sqrt 2} \left( \left| 1_\qq ,
0_\QQ \right>_1 \! + \left| 0_\qq , 1_\QQ \right>_1 \right) \, ,
\nonumber \\
J^{PC} = 2^{++}: & \ & X_2 = \left| 1_\qq , 1_\QQ \right>_2 \, .
\label{eq:SwaveQQ}
\end{eqnarray}
In this work it is especially convenient to employ a basis of states
carrying a unique value of $s_\QQ$ and of $s_\qq$.  These states are
$X_1$ and $X_2$ [as seen in Eqs.~(\ref{eq:SwaveQQ})], and
\begin{eqnarray}
{\tilde X}_0 & \equiv & \left| 0_\qq , 0_\QQ \right>_0 =
+ \frac{1}{2} X_0 + \frac{\sqrt{3}}{2} X_0^\prime \, , \nonumber \\
{\tilde X}_0^\prime & \equiv & \left| 1_\qq , 1_\QQ \right>_0 =
+ \frac{\sqrt{3}}{2} X_0 - \frac{1}{2} X_0^\prime \, , \nonumber \\
{\tilde Z} & \equiv & \left| 1_\qq , 0_\QQ \right>_1 =
\frac{1}{\sqrt{2}} \left( Z^\prime \! + Z \right) \, , \nonumber \\
{\tilde Z}^\prime & \equiv & \left| 0_\qq , 1_\QQ \right>_1 =
\frac{1}{\sqrt{2}} \left( Z^\prime \! - Z \right) \, .
\label{eq:HQbasis}
\end{eqnarray}

Including ($u,d$) light-quark flavor produces 12 states: 6 each
with $I \! = \! 0$ and $I \! = \! 1$, and spin structures in the form
of Eqs.~(\ref{eq:Swavediquark}), (\ref{eq:SwaveQQ}), or
(\ref{eq:HQbasis}).  The basis of Eqs.~(\ref{eq:HQbasis}) in
particular is ideal for discussing SU(3)$_{\rm flavor}$ multiplets: A
state component like $1_{u\bar s}$ is a pure flavor octet that
transforms under spin and flavor analogously to $K^{*+}$ (although in
a diquark model it comprises a mixture of color-singlet and
color-octet components).  The full SU(3)$_{\rm flavor}$ structure of
the multiplet $\Sigma^+_g(1S)$ thus consists of 6 octets and 6
singlets.  A study of the possible mixing of states with the same
$J^P$ between different SU(3)$_{\rm flavor}$ octets, or of
octet-singlet mixing, form two principal theory innovations of this
work. 

The $\QQ q\bar q^\prime$ states in the multiplet $\Sigma^+_g(1S)$
are sufficient to accommodate all particles considered in this work.
However, we note that Ref.~\cite{Lebed:2017min} also provides a
classification of orbitally excited states (the multiplets
$\Sigma^+_g(nP)$ appearing in Ref.~\cite{Giron:2020fvd}), as well as
states in excited-glue Born-Oppenheimer potentials such as
$\Pi_u^+$ (which are exotic analogues to hybrid mesons), and
pentaquark states $\QQ q \hspace{1pt} q_1 q_2$.

\section{Review of $\ccqq$ and $\ccss$ Sector}
\label{sec:Ham}
\subsection{$\ccqq$ Sector}

For hidden heavy-flavor exotics containing only $u$ and/or $d$ light
valence quarks, we write the following Hamiltonian:
\begin{eqnarray} \label{eq:MasterHam}
H &=& M_0 + \Delta H_{\kqQ} + \Delta H_{V_0} + \Delta H_{V_8}
\nonumber\\ &=&
M_0 + 2\kqQ\left(\mathbf{s}_q\cdot \mathbf{s}_Q+ \mathbf{s}_{\bar{q}}
\cdot \mathbf{s}_{\bar{Q}}\right) +
V_0\left(\bm{\tau}_q \! \cdot\bm{\tau}_{\bar{q}}\right)
\left(\bm{\sigma}_{q} \! \cdot\bm{\sigma}_{\bar{q}}\right)
\nonumber\\ &&+ V_8\left(\lambda^8_a \, \bm{\sigma}_q\right)
\! \cdot \! \left(\lambda^8_a \, \bm{\sigma}_{\bar{q}}\right).
\end{eqnarray}
Here, $M_0$ is the common $\Sigma^+_g(1S)$ multiplet mass, which
depends only upon the chosen diquark ($\de , \, \bde$) masses and a
central potential $V(r)$ computed numerically on the lattice from
pure glue configurations that connect $\bf{3}$ and $\bar{\bf{3}}$
sources, as employed in Ref.~\cite{Giron:2019bcs}.  The second term
represents the spin-spin interaction within diquarks, assumed to
couple only $q \leftrightarrow Q$ and $\bar{q}' \leftrightarrow
\bar{Q}$, and $\kqQ$ indicates the strength of this interaction.  The
prime (flavor) index on the light antiquark has been suppressed
throughout Eq.~(\ref{eq:MasterHam}), since for the moment we consider
$q,q^\prime$ to be either a light-quark or strange-quark pair, so
that the same value of $\kappa_{qQ}$ appears for both spin-spin
terms.  An isospin-spin-dependent interaction of strength $V_0$
between the light-quark spins, which is modeled on the pion-nucleon
coupling and was first introduced in Ref.~\cite{Giron:2019cfc},
comprises the third term.  These 3 terms form the full set included
in the analysis of  Ref.~\cite{Giron:2019cfc}.  The final term is new
to this work; it is modeled on an $\eta$-nucleon coupling and
evaluates in the relevant flavor sectors to:
\begin{eqnarray} \label{eq:V8mat}
\Delta M_{V_8}= \frac 1 3 V_8\left[2s_\qq \! \left(s_\qq +1 \right)
-3\right] \times \!
\begin{cases}
1 ,\;q,q^\prime \in \{ u, d \}\\
4, \;q,q^\prime = s, s
\end{cases} \hspace{-1.0em} . \nonumber\\
\end{eqnarray}

To compute the mass expressions arising from
Eq.~(\ref{eq:MasterHam}), let us first abbreviate
\begin{equation}
V_{-} \equiv V_0 - \frac 1 9 V_8 \, , \ \
V_{+} \equiv V_0 + \frac 1 3 V_8 \, .
\end{equation}
Then the Hamiltonian matrix elements for the mixed $Q\bar Q q\bar
q^\prime$ states of the $\Sigma^+_g(1S)$ multiplet, their components
arranged in the order $s_{\QQ} \! = \! 0,1$, read
\begin{eqnarray}
\label{eq:QQqqmix}
\tilde{M}_{0^{++}}^{I=0}&=& M_0
\begin{pmatrix}
1 & 0\\ 
0 & 1
\end{pmatrix}
-\kqQ
\begin{pmatrix}
0 & \sqrt{3}\\
\sqrt{3} & 2
\end{pmatrix}
-3V_{-}
\begin{pmatrix}
-3 & 0\\
0 & 1
\end{pmatrix} \! ,\nonumber\\
\tilde{M}_{0^{++}}^{I=1}&=&M_0
\begin{pmatrix}
1 & 0\\
0 & 1
\end{pmatrix}
-\kqQ
\begin{pmatrix}
0 & \sqrt{3}\\
\sqrt{3} & 2
\end{pmatrix}
+V_+
\begin{pmatrix}
-3 & 0\\
0 & 1
\end{pmatrix},\nonumber\\
\tilde{M}_{1^{+-}}^{I=0}&=&
M_0
\begin{pmatrix}
1 & 0\\
0 & 1
\end{pmatrix}
+\kqQ
\begin{pmatrix}
0 & 1\\
1 & 0
\end{pmatrix}
-3V_-
\begin{pmatrix}
1 & 0\\
0 & -3
\end{pmatrix},\nonumber\\
\tilde{M}_{1^{+-}}^{I=1}&=&
M_0
\begin{pmatrix}
1 & 0\\
0 & 1
\end{pmatrix}
+\kqQ
\begin{pmatrix}
0 & 1\\
1 & 0
\end{pmatrix}
+V_+
\begin{pmatrix}
1 & 0\\
0 & -3
\end{pmatrix}.
\end{eqnarray}
Diagonalizing the expressions of Eq.~(\ref{eq:QQqqmix}) in order of
increasing mass and appending the corresponding (already diagonal)
expressions for the remaining $Q\bar Q q\bar q^\prime$ states of the
$\Sigma^+_g(1S)$ multiplet, one obtains
\begin{eqnarray} \label{eq:MassDiag}
M_{0^{++}}^{I=0}&=&\left(M_0-\kqQ +3V_-\right)
\begin{pmatrix}
1 & 0 \\
0 & 1
\end{pmatrix}+2\tilde{V}_1^{\QQqq}\begin{pmatrix}
-1 & 0\\
0 & 1
\end{pmatrix} , \nonumber\\
M_{0^{++}}^{I=1}&=&\left(M_0 - \kqQ - V_+\right)
\begin{pmatrix}
1 & 0 \\
0 & 1
\end{pmatrix}+2\tilde{V}_2^{\QQqq}\begin{pmatrix}
-1 & 0\\
0 & 1
\end{pmatrix}\, ,\nonumber\\
M_{1^{+-}}^{I=0}&=& \left(M_0 + 3V_-\right)
\begin{pmatrix}
1 & 0 \\
0 & 1
\end{pmatrix}+\tilde{V_3}^{\QQqq}\begin{pmatrix}
-1 & 0 \\
0 & 1
\end{pmatrix}\, , \nonumber\\
M_{1^{+-}}^{I=1}&=&\left(M_0 - V_+\right)
\begin{pmatrix}
1 & 0 \\
0 & 1
\end{pmatrix}+\tilde{V}_4^{\QQqq}\begin{pmatrix}
-1 & 0 \\
0 & 1
\end{pmatrix}\, , \nonumber\\
M_{1^{++}}^{I=0}&=&M_0-\kqQ -3V_- \, ,\nonumber\\
M_{1^{++}}^{I=1}&=&M_0-\kqQ +V_+ \, ,\nonumber\\
M_{2^{++}}^{I=0}&=&M_0+\kqQ -3V_- \, ,\nonumber\\
M_{2^{++}}^{I=1}&=&M_0+\kqQ +V_+ \, ,
\end{eqnarray}
using the abbreviations
\begin{eqnarray}
\tilde{V}_1^{\QQqq} &\equiv & \sqrt{\kqQ^2 +3\kqQ V_- +
9V_{-}^2} \, , \nonumber\\
\tilde{V}_2^{\QQqq} &\equiv & \sqrt{\kqQ^2 -\kqQ V_+ +V_{+}^2}
\, , \nonumber\\
\tilde{V}_3^{\QQqq} &\equiv & \sqrt{\kqQ^2 + 36V_{-}^2} \, ,
\nonumber\\
\tilde{V}_4^{\QQqq} &\equiv & \sqrt{\kqQ^2 + 4V_{+}^2} \, .
\end{eqnarray}

\subsection{$\ccss$ Sector}

The Hamiltonian relevant to the $\ccss$ sector is identical to the
one in Eq.~(\ref{eq:MasterHam}), omitting the isospin-dependent $V_0$
term and performing some $q \! \to \! s$ relabeling,
\begin{eqnarray} \label{eq:ssHam}
H &=& M_0 + \Delta H_{\ksQ} + \Delta H_{V_0} + \Delta H_{V_8}
\nonumber\\ &=&
M_0 + 2\ksQ\left(\mathbf{s}_s\cdot \mathbf{s}_Q+ \mathbf{s}_{\bar{s}}
\cdot \mathbf{s}_{\bar{Q}}\right) +
V_8\left(\lambda^8_a \, \bm{\sigma}_s\right)
\! \cdot \! \left(\lambda^8_a \, \bm{\sigma}_{\bar{s}}\right) \, .
\nonumber \\
\end{eqnarray}
Equivalently, this expression generalizes the Hamiltonian used in the
analysis of Ref.~\cite{Giron:2020qpb} by the inclusion of the $V_8$
term.  Note that the value of $M_0 \! = \! M_0^{\QQss}$ here differs
from $M_0 \! = \! M_0^{\QQqq}$ appearing in Eq.~(\ref{eq:MasterHam}).
The mass eigenvalues for the 6 isosinglet $c\bar c s\bar s$ states
evaluate to
\begin{eqnarray} \label{eq:ccssMass}
M_{0^{++}}&=&\left(M_0 -\ksQ-\frac{4}{3}V_8\right)\begin{pmatrix}
1 & 0 \\
0 & 1
\end{pmatrix}
\! +2\tilde{V}_1^{\QQss} \! \begin{pmatrix}
-1 & 0 \\
0 & 1
\end{pmatrix} \! ,\nonumber\\
M_{1^{+-}}&=& \left(M_0-\frac{4}{3}V_8\right)
\begin{pmatrix}
1 & 0 \\
0 & 1
\end{pmatrix}
+ \tilde{V}_2^{\QQss}
\begin{pmatrix}
-1 & 0 \\
0 & 1
\end{pmatrix},\nonumber\\
M_{1^{++}} &=& M_0 -\ksQ + \frac{4}{3}V_8 \, ,\nonumber\\
M_{2^{++}} &=& M_0 +\ksQ + \frac{4}{3}V_8 \, ,
\end{eqnarray}
where
\begin{eqnarray}
\tilde{V}_1^{\QQss} &\equiv & \sqrt{\ksQ^2 -\frac{4}{3}\ksQ V_8 +
\frac{16}{9}V_8^2} \, , \nonumber\\
\tilde{V}_2^{\QQss} &\equiv & \sqrt{\ksQ^2 + \frac{64}{9} V_8^2} \, .
\end{eqnarray}

\section{Hidden-Charm, Open-Strange Sector}
\label{sec:OpenStrange}

In this sector, the Hamiltonian analogous to Eq.~(\ref{eq:MasterHam})
becomes
\begin{eqnarray} \label{eq:StrangeHam}
H &=& H_0 +\Delta H_{\kqQ} + \Delta H_{\kappa_{s Q}}+ \Delta H_{V_8}
\nonumber\\
&=& H_0 + 2\left[\kqQ\left(\bold{s}_q \cdot \bm{s}_Q \right)+
\kappa_{sQ} \left(\bold{s}_{\bar{s}} \cdot \bm{s}_{\bar{Q}}\right)
\right] \nonumber \\
&& + V_8\left(\lambda^8_a \, \bm{\sigma}_q \right) \! \cdot \!
\left(\lambda^8_a \, \bm{\sigma}_{\bar{s}}\right) \, .
\end{eqnarray}
Without loss of generality, we have taken $q^\prime \! \to \! s$,
with the opposite choice $q \! \to \! s$ simply leading to the
antiparticles of those studied here.  Then the spin couplings
$\kappa_{sQ}$ and $\kappa_{qQ}$ are numerically quite distinct, and
we compute the mass contributions
\begin{eqnarray}
\Delta M_{\kqQ} =\frac 1 2 \kqQ\left[2s_\de\left(s_\de +1\right)-3
\right] \, ,
\end{eqnarray}
\begin{eqnarray}
\Delta M_{\kappa_{\bar s \bar Q}}= \frac 1 2 \kappa_{sQ}
\left[2s_\bde \left(s_\bde +1\right) -3 \right] \, ,
\end{eqnarray}
and
\begin{eqnarray}
\Delta M_{V_8} = -\frac{2}{3}V_8\left[2s_{q\bar s}\left( s_{q\bar s}
+1\right) -3\right] \, ,
\end{eqnarray}
with the final expression computed in the same manner as is
performed to obtain Eq.~(\ref{eq:V8mat}).

A notable feature of the open-strange exotics sector becomes apparent
when considering the full SU(3)$_{\rm flavor}$ multiplet structure.
$Q\bar Q q\bar q$ states with $I_3 \! = \! 0$ carry good $J^{PC}$
quantum numbers, and states with different $J^{PC}$ values of course
cannot mix with them.  Inasmuch as isospin is a nearly exact
symmetry, one can extend $C$ parity to a full isospin multiplet by
defining the conserved $G$-parity quantum number (whose eigenvalues
for all hadrons are tabulated by the Particle Data Group
(PDG)~\cite{Zyla:2020zbs}).  Specifically,
\begin{equation}
G \equiv (-1)^I C \, ,
\end{equation}
where the $C$-parity eigenvalue here is that of the $I_3 \! = \! 0$
member of the isomultiplet.  One could generalize the concept of $G$
parity to a full SU(3)$_{\rm flavor}$ multiplet, but since the
corresponding flavor symmetry is broken, mixing between the
open-strange members of multiplets whose $I_3 \! = \! 0$,
$Y \! = \! 0$ members have opposite $C$ parities can occur.  Indeed,
this phenomenon is known among the conventional mesons: For example,
the strange partners to the lightest $1^{++}$ and $1^{+-}$ mesons are
named $K_{1A}$ and $K_{1B}$ respectively, and the observed $1^+$
strange-meson mass eigenstates $K_1(1270)$ and $K_1(1400)$ are
believed to be nearly equal admixtures of $K_{1A}$ and
$K_{1B}$~\cite{Zyla:2020zbs}.

In the exotics sector, the nonstrange $X_1$ ($1^{++}$) states cannot
mix with $\tilde{Z}, \tilde{Z}^\prime$ ($1^{+-}$) due to $G$-parity
conservation.  However, their open-strange $1^+$ partners {\em can\/}
mix, leading to richer phenomenological possibilities.  To wit: The
mass expressions obtained from Eq.~(\ref{eq:StrangeHam}), prior to
diagonalization, read
\begin{eqnarray} \label{eq:OpenStrangeExact}
\tilde{M}_{0^{+}}&=& M_0 
\begin{pmatrix}
1 & 0 \\
0 & 1 \\
\end{pmatrix}
-\frac{1}{2} \left( \kqQ + \kappa_{sQ} \right)
\begin{pmatrix}
0 & \sqrt{3}\\
\sqrt{3} & 2
\end{pmatrix}
\nonumber \\ & &
-\frac{2}{3}V_8
\begin{pmatrix}
-3 & 0\\
0 & 1
\end{pmatrix} \, , \nonumber\\
\tilde{M}_{1^+}&=& M_0
\begin{pmatrix}
1 & 0 & 0\\
0 & 1 & 0\\
0 & 0 & 1
\end{pmatrix} + \frac{\kqQ}{2}
\begin{pmatrix}
-1 & -\sqrt{2} & +\sqrt{2} \\
-\sqrt{2} & 0 & +1  \\
+\sqrt{2} & +1  & 0 
\end{pmatrix} \, , \nonumber\\
&&+ \frac{\kappa_{sQ}}{2}
\begin{pmatrix}
-1 & +\sqrt{2} & -\sqrt{2} \\
+\sqrt{2} & 0 & +1 \\
-\sqrt{2} & +1 & 0
\end{pmatrix}
-\frac{2}{3}V_8
\begin{pmatrix}
1 & 0 & 0\\
0 & -3 & 0 \\
0 & 0 & 1 \\
\end{pmatrix} \, , \nonumber\\
M_{2^{+}}&=& M_0+\frac{1}{2}\left(\kqQ + \kappa_{sQ} \right)
-\frac{2}{3}V_8 \, .
\end{eqnarray}
The elements of the matrices for $0^+$ are again arranged in order of
increasing heavy-quark spin.  However, those for $1^+$ are arranged
in the order corresponding to increasing mass eigenvalues for their
nonstrange partners in the hidden-charm sector:
$X_1, {\tilde Z}^\prime, {\tilde Z}$.

The mass eigenvalues for the $0^+$ sector read
\begin{eqnarray}
M_{0^+}&=&\left[M_0 -\frac 1 2 \left( \kqQ + \kappa_{sQ} \right)
+\frac{2}{3}V_8\right]\begin{pmatrix}
1 & 0 \\
0 & 1
\end{pmatrix}
\nonumber\\ & &
+2\tilde{V}_1^{\QQqs}\begin{pmatrix}
-1 & 0 \\
0 & 1
\end{pmatrix} ,
\end{eqnarray}
where
\begin{equation}
\tilde{V}_1^{\QQqs} \equiv \sqrt{ \left[ \frac{1}{2} \left( \kqQ +
\kappa_{sQ} \right) + \frac{1}{3} V_8 \right]^2 + \frac{1}{3} V_8^2 }
\, .
\end{equation}
The exact expressions for the $1^+$ eigenvalues are of course
complicated roots of a cubic equation, but anticipating that
$V_8 \! \ll \! \kqQ \! \ll \! \kappa_{sQ}$, one may perform a
perturbative expansion in $V_8$ to compute approximate values:
\begin{eqnarray} \label{eq:ZcsApprox}
M_{1^+}^{(1)} & = & M_0^{\QQqs} + \frac{1}{2}
\left( -3\kappa_{sQ} +\kqQ \right) + O(V_8^2/\kappa_{sQ}) \, ,
\nonumber \\
M_{1^+}^{(2)} & = & M_0^{\QQqs} + \frac{1}{2}
\left( \kappa_{sQ} -3\kqQ \right) + O(V_8^2/\kqQ) \, , \nonumber \\
M_{1^+}^{(3)} & = & M_0^{\QQqs} + \frac{1}{2}
\left( \kappa_{sQ} +\kqQ \right) +\frac{2}{3} V_8 + O(V_8^2/\kqQ ) \,
. \nonumber \\
\end{eqnarray}

\section{Analysis}
\label{sec:Analysis}

\subsection{Flavor SU(3) Multiplets and Mixing}
\label{subsec:SU3Mixing}

The original analysis of $c\bar c s\bar s$ states in
Ref.~\cite{Lebed:2016yvr}, as well as its updated form in
Ref.~\cite{Giron:2020qpb}, takes the $c\bar c s\bar s$ states to be
completely unmixed with those in the $c\bar c q\bar q^\prime$ sector,
where $q, q^\prime \! \in \{ u, d \}$.  If, on the other hand,
SU(3)$_{\rm flavor}$ is exact, then the states should fill octets and
singlets of the flavor symmetry.  Specifically, the flavor structure
of $c\bar c q\bar q^\prime$ states, now allowing
$q, q^\prime \! \in \{ u, d, s \}$, can be discussed using the same
framework that applies to conventional $q\bar q^\prime$ mesons.  The
$I \! = \! 0$, $I_3 \! = \! 0$, $Y \! = \! 0$ unmixed octet and
singlet combinations are, as usual,
\begin{equation}
\frac{1}{\sqrt{6}} \left( u\bar u + d\bar d - 2s\bar s \right)
\ {\rm and} \
\frac{1}{\sqrt{3}} \left( u\bar u + d\bar d + s\bar s \right) \, ,
\end{equation}
respectively.  In the lightest ($J^{PC} \! = \! 0^{-+}$) meson
multiplet, these states correspond to $\eta$ and $\eta^\prime$,
respectively, which remain largely unmixed because the octet states
are pseudo-Nambu-Goldstone bosons whose masses vanish in the chiral
limit, while the singlet has a nonzero mass in this limit due to the
anomalous breaking of the axial U(1) symmetry of massless QCD\@.

Heavier meson multiplets, however, support much larger
SU(3)$_{\rm flavor}$ mixing between the octet and singlet
combinations.  For example, the next-lightest ($1^{--}$) multiplet
features the $\omega$ and $\phi$ as its $I \! = \! 0$ states, which
appear to be nearly ideally mixed into the flavor combinations
$\frac{1}{\sqrt{2}} ( u\bar u + d\bar d )$ and $s\bar s$,
respectively.  The appearance of only the $J/\psi \, \phi$ decay mode
for most of the purported $c\bar c s\bar s$ candidates inspired the
implicit adoption of an ideal-mixing ansatz in
Refs.~\cite{Lebed:2016yvr,Giron:2020qpb}.

Moreover, the approach of treating $I \! = \! 0$ states in the
$c\bar c q\bar q^\prime$ sector as containing no $s\bar s$ component,
which is implicit in Refs.~\cite{Giron:2019bcs,Giron:2019cfc,
Giron:2020qpb}, also introduces a hidden assumption of octet-singlet
mixing into the analysis.  The fact that $\rho^0$ ($I \! = \! 1$,
pure octet) and $\omega$ ($I \! = \! 0$, ideally mixed) are nearly
degenerate in mass, and likewise for $Z_c(3900)^0$ ($I \! = \! 1$,
pure octet) and $X(3872)$ ($I \! = \! 0$), suggests that substantial
octet-singlet flavor mixing is needed to understand the spectrum of
both $1^{--}$ conventional mesons and $\Sigma_g^+(1S)$ hidden-charm
exotic mesons.  Isospin symmetry then links the remaining
$Y \! = \! 0$ states ($\rho^{\, \pm}$, $Z_c(3900)^\pm$).  In the case
of exotics, the values of $M_0^{c\bar c q\bar q^\prime}$ and
$M_0^{c\bar c s\bar s}$ extracted in previous work implicitly
incorporate ideal octet-singlet flavor mixing, with strangeness
dependence entering only through the differing diquark masses for
$\de \! = \! (cq)$ and for $\de \! = \! (cs)$.

In contrast, the value of $M_0^{c\bar c q\bar s}$ for the
open-strange ($K$-like) states refers to an unmixed
SU(3)$_{\rm flavor}$ octet.  One should therefore not be surprised
that simply combining the values of $m_{(cq)}$ from
$M_0^{c\bar c q\bar q^\prime}$ and $m_{(cs)}$ from
$M_0^{c\bar c s\bar s}$ with the lattice-computed glue potential
$V(r)$ between heavy color-triplet sources generates a value for
$M_0^{c\bar c q\bar s}$ slightly different from the one that would be
obtained from starting with values of $M_0$ corresponding to unmixed
SU(3)-octet or -singlet $c\bar c q\bar q^\prime$ and
$c\bar c s\bar s$ states.  This effect should occur even in the
absence of an explicit SU(3)$_{\rm flavor}$-breaking difference
between $m_{\de = (cq)}$ and $m_{\de = (cs)}$.

We illustrate this mixing effect using a toy example well known from
elementary quantum mechanics: Ignore fine-structure effects and let
the unmixed mass parameters $M_8$ (pure octet) and $M_1$ (pure
singlet) be degenerate, $M \! = \! M_8 \! = \! M_1$, in a 2-level
system with an octet-singlet mass-mixing parameter $\Delta$.  Then
the resulting mass eigenvalues are $M \mp \Delta$, and the mixing
angle of the system is maximal, $45^\circ$.  More generally, the
lower mass eigenvalue is always smaller than the smaller diagonal
element, whether or not the unmixed octet and singlet mass parameters 
are equal.  In our case, the value of $M_0^{c\bar c q\bar q^\prime}$
from the previous analyses of Refs.~\cite{Giron:2019bcs,
Giron:2019cfc,Lebed:2016yvr,Giron:2020qpb} is assumed to refer to
ideally mixed states; and since one expects the exotics observed thus
far (which are used to extract $M_0$ values) to be the lightest ones
that exist, the derived $M_0^{c\bar c q\bar q^\prime}$ results
represent the smaller mass eigenvalues.  Meanwhile,
$M_0^{c\bar c s\bar s}$ is extracted from states assumed to contain
no light valence quarks, and therefore their component diquarks are
pure $(cs)$; thus, no mixing needs to be performed to extract the
parameter $m_{\de = (cs)}$.  The result for $M_0^{c\bar c q\bar s}$
naively obtained from using the (mixed)
$M_0^{c\bar c q\bar q^\prime}$ and $M_0^{c\bar c s\bar s}$ values
should therefore be slightly lower than one determined entirely from
the pure-octet $c\bar c q\bar s$ sector.  This expectation, in fact,
is precisely what occurs, as we see below.

\subsection{$c\bar c q\bar q^\prime$ Sector and $V_8$}

The analysis of the $c\bar c q\bar q^\prime$ sector here closely
follows that of Ref.~\cite{Giron:2019cfc}, and especially
Ref.~\cite{Giron:2020qpb}.  The 3 primary inputs are the PDG
averages~\cite{Zyla:2020zbs}
\begin{eqnarray}
m_{X(3872)}   & = & 3871.69 \pm 0.17 \ {\rm MeV} \, , \nonumber \\
m_{Z_c(3900)} & = & 3888.4 \pm 2.5   \ {\rm MeV} \, , \nonumber \\
m_{Z_c(4020)} & = & 4024.1 \pm 1.9   \ {\rm MeV}  \, ,
\label{eq:Masses}
\end{eqnarray}
with only the value for $Z_c(3900)$ changing slightly since the
previous analyses.  Since the Hamiltonian of Eq.~(\ref{eq:MasterHam})
now has 4 parameters, the system is underdetermined.  However, one
further constraint arises from noting the strong charmonium decay
preference~\cite{Zyla:2020zbs} of $Z_c(3900)$ to $J/\psi$, and
$Z_c(4020)$ to $h_c$, suggesting that these $Z_c$ states are nearly
pure $s_{c\bar c} \! = \! 1$ and $s_{c\bar c} \! = \! 0$ eigenstates,
respectively.  Defining $P$ as the $s_{Q\bar Q} \! = \! 1$
probability content of the lower-mass $1^{+-}$, $I \! = \! 1$
eigenstate of Eqs.~(\ref{eq:MassDiag}) [{\it i.e.}, the square of the
off-diagonal component of the unitary matrix diagonalizing
${\tilde M}^{I = 1}_{1^{+-}}$ in Eqs.~(\ref{eq:QQqqmix})], one
obtains
\begin{equation}
P = \frac 1 2 \left[ 1 + \frac{2 \left( V_0 + \frac{V_8}{3} \right)}
{\sqrt{\kqQ^2 + 4 \left( V_0 + \frac{V_8}{3} \right)^2}} \right] \, ,
\end{equation}
which means that $V_8$ can be expressed as a function of $P$ (and the
parameters $V_0$ and $\kqQ$).  Using this constraint with the mass
expressions in Eqs.~(\ref{eq:MassDiag}), the most convenient
combinations of the 3 masses in Eqs.~(\ref{eq:Masses}) are
\begin{eqnarray}
\mu_1 & \equiv & \frac 1 2 \left( m_{Z_c(4020)} + m_{Z_c(3900)}
\right) = 3956.3 \pm 1.6 \ {\rm MeV} \nonumber \\
& = & M_0 - \frac 1 2 \left( P - \frac 1 2 \right)
\frac{\kqQc}{\sqrt{P(1-P)}} \, , \nonumber \\
\mu_2 & \equiv & \frac 1 2 \left( m_{Z_c(4020)} + m_{Z_c(3900)}
\right) - m_{X(3872)} \nonumber \\
& = & 84.6 \pm 1.6 \ {\rm MeV} \nonumber \\
& = & \kqQc \left[  1 - \frac{\left( P - \frac 1 2 \right)}
{\sqrt{P(1-P)}} \right] + 4V_0 \, , \nonumber \\
\mu_3 & \equiv & \frac 1 2 \left( m_{Z_c(4020)} - m_{Z_c(3900)}
\right) = 67.9 \pm 1.6 \ {\rm MeV} \nonumber \\
& = & \frac{\kqQc}{2\sqrt{P(1-P)}} \, .
\label{eq:Nums}
\end{eqnarray}
From Eqs.~(\ref{eq:Nums}), one extracts
\begin{equation} \label{eq:V8inP}
\frac 4 3 V_8 = -\mu_2 +2\mu_3 \left[ \left( P - \frac 1 2 \right)
+ \sqrt{P(1-P)} \right] \, .
\end{equation}
The case $V_8 \! = \! 0$, which (in effect) is imposed in
Ref.~\cite{Giron:2020qpb}, becomes
\begin{eqnarray} \label{eq:Pvalues}
P & = & \frac 1 4 \left[ 2 + \frac{\mu_2}{\mu_3} \pm
\sqrt{ 2 - \left(\frac{\mu_2}{\mu_3}\right)^2} \, \right] \, ,
\nonumber \\
& = & 0.979 \pm 0.009 \, , \ 0.644 \pm 0.030 \, .
\end{eqnarray}
The only input used in Ref.~\cite{Giron:2020qpb} beyond those of
Eqs.~(\ref{eq:Masses}) is the discrete choice of the larger $P$ value
in Eq.~(\ref{eq:Pvalues}) to recognize the $Z_c$ charmonium decay
preferences noted above.

In fact, Eq.~(\ref{eq:V8inP}) places a rather strong constraint upon
$V_8$.  While any $P \! \in \! [0,1]$ is in principle allowed, values
of $P$ smaller (larger) than the smaller (larger) root in
Eq.~(\ref{eq:Pvalues}) lead to negative values of $V_8$---and, for
sufficiently small values of $P$, values of $V_8$ that are also
larger in magnitude than $V_0$.  Inasmuch as the accompanying
operators in Eq.~(\ref{eq:MasterHam}) represent $\pi$-like and
$\eta$-like exchanges, respectively, one expects the analogy to the
dynamics of true $\pi$ and $\eta$ exchanges between nucleons (from,
{\it e.g.}, chiral perturbation theory) to hold.  Under this
assumption, the $\eta$-like exchange should be attractive like the
$\pi$-like exchange; hence $V_8$, like $V_0$, should be positive.
However, genuine $\eta$ exchange is also weaker than $\pi$ exchange,
both due to the $\eta$'s larger mass and larger decay constant.  We
therefore take $V_8 \! > \! 0$ to be a natural constraint of the
model, which requires $P$ to lie between the roots given in
Eq.~(\ref{eq:Pvalues}).  According to Eq.~(\ref{eq:V8inP}), within
this range $V_8$ reaches a maximum at $P = \! \frac 1 2 \big( 1 +
\frac{1}{\sqrt{2}} \big) \! = \! 0.854$, at which
\begin{equation} \label{eq:V8max}
V_8^{\rm max} = \frac 3 4 \left( -\mu_2 + \sqrt{2} \mu_3 \right) =
8.6 \pm 2.3 \ {\rm MeV} \, .
\end{equation}

We therefore expect the allowed range of $V_8$, as determined by the
known phenomenology of the $c\bar c q\bar q^\prime$ sector, to have a
modest effect compared to that provided by the other parameters in
Eq.~(\ref{eq:MasterHam}).  Using Eqs.~(\ref{eq:Nums}), one obtains
for the other Hamiltonian parameters:
\begin{eqnarray} \label{eq:ccqqParams}
M_0 & = & \mu_1 + \left( P - \frac 1 2 \right) \mu_3 \, , \nonumber
\\
\kappa_{qc} & = & 2\mu_3 \sqrt{P(1-P)} \, , \nonumber \\
4V_0 & = & \mu_2 + 2\mu_3 \left[ \left( P - \frac 1 2 \right) -
\sqrt{P(1-P)} \right] \, .
\end{eqnarray}

The values of $M_0$, $\kqQc$, and $V_0$ obtained for both
$V_8 \! = \! 0$ and for an optimized $V_8$ value obtained below from
the $c\bar c s\bar s$ spectrum [in Eqs.~(\ref{eq:ccssBest})] are
presented in Table~\ref{tab:ModelParams}.  The full spectrum of
masses for the $c\bar c q\bar q^\prime$ $\Sigma^+_g(1S)$ multiplet
appears in Table~\ref{tab:MassSpectrumAll}\@.

\begin{table}
\caption{Hamiltonian parameters [Eq.~(\ref{eq:MasterHam})] of the
dynamical diquark model obtained from fits to members of the
$\Sigma^+_g(1S)$ multiplet in the $c\bar c q\bar q^\prime$
[$X(3872)$, $Z_c(3900)$, $Z_c(4020)$] and $c\bar c s\bar s$
[$X(3915)$, $X(4140)$, $X(4350)$] sectors.  Also included is the
$s_{c\bar c} \! = \! 1$ content $P$ of the state $Z_c(3900)$, and
the diquark mass $m_\delta$ derived from each case, as well as a
value of $M_0^{c\bar c q\bar s}$ for the open-strange sector that
supposes no SU(3)$_{\rm flavor}$ octet-singlet mixing in the other
flavor sectors.}
\centering
\setlength{\extrarowheight}{1.2ex}
\begin{tabular}{ c | c | c }
\hline\hline
& $V_8 = 0$ & $\ V_8 = 3.9 \pm 1.4 \ {\rm MeV}$ \\
\hline
$M_0^{c\bar cq\bar q^\prime}$ & $3988.7 \pm 1.5 \ {\rm MeV}$
& $3987.7 \pm 1.6 \ {\rm MeV}$ \\
$\kappa_{qc}$ & $19.6 \pm 3.7 \ {\rm MeV}$
& $25.6 \pm 5.0 \ {\rm MeV}$ \\
$V_0$ & $32.5 \pm 1.3 \ {\rm MeV}$ & $30.4 \pm 1.9 \ {\rm MeV}$ \\
$P$ & $0.979 \pm 0.009$ & $0.963 \pm 0.016$ \\
$m_{\de (cq)}$ & $1927.0 \pm 11.5 \ {\rm MeV}$
& $1927.1 \pm 11.0 \ {\rm MeV}$ \\ 
\hline
$M_0^{c\bar c s\bar s}$ & $\cdots$ & $4251.3 \pm 2.8 \ {\rm MeV}$ \\
$\kappa_{sc}$ & $\cdots$ & $109.8 \pm 1.1 \ {\rm MeV}$ \\
$m_{\de (cs)}$ & $\cdots$ & $2069.4 \pm 10.9 \ {\rm MeV}$ \\
\hline
$M_0^{c\bar c q\bar s}$ & $\cdots$ & $4119.7 \pm 1.7\ {\rm MeV}$ \\
\hline
\end{tabular}
\label{tab:ModelParams}
\end{table}

Using the value of $M_0^{c\bar cq\bar q^\prime}$ from
Table~\ref{tab:ModelParams} with $V_8 \! = \! 0$ and the
lattice-simulated glue potentials $V(r)$ of Refs.~\cite{Juge:2002br,
Morningstar:2019} (JKM) and \cite{Capitani:2018rox} (CPRRW), we
compute
\begin{eqnarray} \label{eq:cqDiquarkNoV8}
m_{\de(cq)} & = & 1938.5 \pm 0.8 \ {\rm MeV} \ {\rm (JKM)} \, ,
\nonumber \\
& = & 1915.5 \pm 0.8 \ {\rm MeV} \ {\rm (CPRRW)} \, .
\end{eqnarray}
A value spanning this spread is presented in
Table~\ref{tab:ModelParams}.  Again, these results
for $V_8 \! = \! 0$ differ from those in Ref.~\cite{Giron:2020qpb}
only through a small shift in the tabulated PDG value of
$m_{Z_c(3900)}$ in Eq.~(\ref{eq:Masses}).

\subsection{$c\bar c s\bar s$ Sector: $\kqQs$ and $V_8$}

The signature process used in Ref.~\cite{Giron:2020qpb} to identify
$c\bar c s\bar s$ exotics is the decay mode $J/\psi \, \phi$,
although some candidates are identifiable through $D_s$-type
meson-pair decays.  Furthermore, the analysis of
Ref.~\cite{Giron:2020qpb} argues that the $J^{PC} \! = \! 1^{++}$
$X(4274)$ is an excellent candidate for the conventional charmonium
state $\chi_{c1}(3P)$.  The most unexpected addition to the
$c\bar c s\bar s$ spectrum is the peculiar state $X(3915)$, with
likely $0^{++}$ quantum numbers, as the candidate for the lightest
$c\bar c s\bar s$ state in this model.  In brief summary of the
reasoning in Ref.~\cite{Lebed:2016yvr} and references therein,
$X(3915)$ has no confirmed open-charm decays, thus arguing against it
being either the conventional charmonium state $\chi_{c0}(2P)$ or
$c \bar c q\bar q$.  It has definitively been seen to couple only to
$\gamma \gamma$ and $J/\psi \, \omega$; with respect to the latter
mode, note that $X(3915)$ lies below the $J/\psi \, \phi$ threshold,
so that $\phi \! \to \! \omega$ mixing is proposed in
Ref.~\cite{Lebed:2016yvr} to be responsible for the
$J/\psi \, \omega$ decay mode.  Furthermore, a recent lattice
calculation~\cite{Prelovsek:2020eiw} predicts the existence of a
$0^{++}$ state in this mass region that has a strong coupling to
$D_s \bar D_s$ but a weak coupling to $D \bar D$.  The mass used in
this work is the PDG value~\cite{Zyla:2020zbs}:
\begin{equation} \label{eq:X3915}
m_{X(3915)} = 3921.7 \pm 1.8 \ {\rm MeV} \, .
\end{equation}

In the previous analysis~\cite{Giron:2020qpb}, the $c\bar c s\bar s$
spectrum obtained for the multiplet is very simple.  Referring to
Eqs.~(\ref{eq:ccssMass}), the assumption that
$\kappa_{sc} \! \gg \! V_8 \! > \! 0$ leads to the spectrum (in
increasing order of mass):
\begin{eqnarray} \label{eq:ccssApprox}
M_{0^{++}} & = & M_0 - 3\kappa_{sc} + O(V_8^2/\kappa_{sc}) \, ,
\nonumber \\
M_{1^{+-}} & = & M_0 -\kappa_{sc} -\frac 4 3 V_8 +
O(V_8^2/\kappa_{sc}) \, , \nonumber \\
M_{1^{++}} & = &  M_0 -\kappa_{sc} +\frac 4 3 V_8 \, , \nonumber \\
M^\prime_{0^{++}} & = & M_0 +\kappa_{sc} -\frac 8 3 V_8 +
O(V_8^2/\kappa_{sc}) \, , \nonumber \\
M^\prime_{1^{+-}} & = & M_0 +\kappa_{sc} -\frac 4 3 V_8 +
O(V_8^2/\kappa_{sc}) \, , \nonumber \\
M_{2^{++}} & = &  M_0 +\kappa_{sc} +\frac 4 3 V_8 \, ,
\end{eqnarray}
which reduces to 3 degenerate sets in the case $V_8 \! = \! 0$, as
listed in Ref.~\cite{Giron:2020qpb}.  In particular, the lighter
$0^{++}$ state clearly lies far below the others, with the $1^{++}$
state (and the lighter $1^{+-}$) being intermediate in mass, and the
$2^{++}$ and heavier $0^{++}$ (and $1^{+-}$) states lying close
together at a larger mass value.  The overall effect of
Eq.~(\ref{eq:ccssApprox}) is to split the $\Sigma^+_g(1S)$ multiplet
into 3 roughly equally spaced (by $2\kappa_{sc}$) clusters of
$c\bar c s\bar s$ states.

The state $X(4140)$ is taken to be an unmistakable $c\bar c s\bar s$
candidate, the sole $1^{++}$ member of the multiplet
$\Sigma^+_g(1S)$.  Therefore, the $c\bar c s\bar s$ spectrum should
start with $X(3915)$ being the distinct lightest member, a $1^{+-}$
state is predicted to appear with a mass near $m_{X(4140)}$, and a
trio of states ($0^{++}$, $1^{+-}$, $2^{++}$) is predicted to appear
at approximately
$m_{X(3915)} \! + \! 2(m_{X(4140)} \! - \! m_{X(3915)})$.  A
complication arises, however, with the latest LHCb
measurement~\cite{Aaij:2021ivw} of $X(4140)$:
\begin{eqnarray} \label{eq:NewX4140}
m_{X(4140)} & = & 4118 \pm 11^{+19}_{-36} \ {\rm MeV} , \nonumber \\
\Gamma_{X(4140)} & = & 162 \pm 21^{+24}_{-49} \ {\rm MeV} \, ,
\end{eqnarray}
which should be compared to the PDG average~\cite{Zyla:2020zbs},
\begin{eqnarray} \label{eq:OldX4140}
m_{X(4140)} & = & 4146.8 \pm 2.4 \ {\rm MeV} , \nonumber \\
\Gamma_{X(4140)} & = & 22^{+8}_{-7} \ {\rm MeV} \, ,
\end{eqnarray}
the mass differing by about $1.3 \, \sigma$ (and the width differing
radically).  LHCb observes $X(4140)$ with a $13 \, \sigma$ total
significance.  On the other hand, the previous LHCb observation of
$X(4140)$~\cite{Aaij:2016iza} (at a significance of $8.4 \, \sigma$)
forms part of the PDG averages of Eqs.~(\ref{eq:OldX4140}), and the
mass value obtained in Ref.~\cite{Aaij:2016iza} is much more in line
with the average mass value given in Eqs.~(\ref{eq:OldX4140}):
\begin{eqnarray} \label{eq:OldLHCb}
m_{X(4140)} & = & 4146.5 \pm 4.5^{+4.6}_{-2.8} \ {\rm MeV} \, ,
\nonumber \\
\Gamma_{X(4140)} & = & 83 \pm 21^{+21}_{-14} \ {\rm MeV} \, .
\end{eqnarray}
In fact, the data used in Ref.~\cite{Aaij:2016iza} forms a small
subset of the LHCb data reported in Ref.~\cite{Aaij:2021ivw}.  So how
can a measurement using much more data lead to a result with much
larger uncertainties?  In large part, it arises from a new modeling
of the $X(4140)$ lineshape, in which a naive Breit-Wigner profile is
replaced with a Flatt{\'e} form~\cite{Tomasz:2021}.  To incorporate
this new development, we reanalyze the PDG mass average of
Eqs.~(\ref{eq:OldX4140}) by replacing the old LHCb mass measurement
of Eqs.~(\ref{eq:OldLHCb}) with the new one of
Eqs.~(\ref{eq:NewX4140}), producing the value to be used in our
analysis:
\begin{equation} \label{eq:FinalX4140}
m_{X(4140)} = 4146.7 \pm 2.7 \ {\rm MeV} \, .
\end{equation}

The state $X(4350)$, although not yet confirmed at the same level of
confidence ($3.2 \sigma$), is seen in
$\gamma\gamma \! \to \! J/\psi \, \phi$ and thus is an excellent
$c\bar c s\bar s$ $0^{++}$ or $2^{++}$ candidate.  Noting
that~\cite{Zyla:2020zbs}
\begin{equation} \label{eq:X4350}
m_{X(4350)} = 4351 \pm 5 \ {\rm MeV} \, ,
\end{equation}
and using Eqs.~(\ref{eq:X3915}) and (\ref{eq:FinalX4140}), one finds
\begin{equation}
m_{X(3915)} + 2(m_{X(4140)} - m_{X(3915)}) = 4371.7 \pm 5.7 \
{\rm MeV} \, .
\end{equation}
One then sees (as in Refs.~\cite{Lebed:2016yvr,Giron:2020qpb}) that
$X(4350)$ nearly satisifies the equal-spacing rule discussed above,
which confirms our previous result that $V_8$ is numerically small.
In fact, at linear order in $V_8$, Eqs.~(\ref{eq:ccssApprox}) give
\begin{eqnarray}
(M_{0^{++}}^\prime - M_{1^{++}}) - (M_{1^{++}} - M_{0^{++}}) & = &
-\frac{16}{3} V_8 \, , \nonumber \\
(M_{2^{++}} - M_{1^{++}}) - (M_{1^{++}} - M_{0^{++}}) & = &
-\frac{4}{3} V_8 \, .
\end{eqnarray}
Using $m_{X(4350)}$ from Eq.~(\ref{eq:X4350}) for $M_{0^{++}}^\prime$
or $M_{2^{++}}$ gives
\begin{eqnarray}
{(m_{X(4350)} - m_{X(4140)}) - (m_{X(4140)} - m_{X(3915)})}
\hspace{2em}
\nonumber \\ = -20.7 \pm 7.6 \ {\rm MeV} \, , \hspace{2em}
\end{eqnarray}
meaning that $V_8$ is small and positive, as anticipated in
Eq.~(\ref{eq:V8max}).  Returning to the full mass expressions of
Eqs.~(\ref{eq:ccssMass}), one uses Eqs.~(\ref{eq:X3915}),
(\ref{eq:FinalX4140}), and (\ref{eq:X4350}) to obtain
\begin{eqnarray} \label{eq:ccssBest}
M_0^{c\bar c s\bar s} & = & 4251.3 \pm 2.8 \ {\rm MeV} \, ,
\nonumber \\
\kappa_{sc} & = & \ \, 109.8 \pm 1.1 \ {\rm MeV} \, , \nonumber \\
V_8 & = & \ \ \ \ \,  3.9 \pm 1.4 \ {\rm MeV} \, ,
\end{eqnarray}
assuming that $X(4350)$ is $0^{++}$, and
\begin{eqnarray}
M_0^{c\bar c s\bar s} & = & 4230.8 \pm 7.0 \ {\rm MeV} \, ,
\nonumber \\
\kappa_{sc} & = & \ \, 102.2 \pm 2.8 \ {\rm MeV} \, , \nonumber \\
V_8 & = & \ \ \; 13.6 \pm 4.3 \ {\rm MeV} \, ,
\end{eqnarray}
assuming that $X(4350)$ is $2^{++}$.  Obtaining these results
requires the resolution of a discrete ambiguity to impose the
physical expectation $\kappa_{sc} \! > \! 0$, as discussed in
Ref.~\cite{Giron:2020qpb}.  The latter solution produces a slightly
larger value of $V_8$ than allowed by Eq.~(\ref{eq:V8max}), but only
by $1.0 \, \sigma$, and therefore still viable.  Nevertheless, for
purposes of illustration, we choose Eqs.~(\ref{eq:ccssBest}) as the
best-fit parameters (also included in Table~\ref{tab:ModelParams}),
and use them to compute the full spectrum of masses for the
$\Sigma^+_g(1S)$ $c\bar c s\bar s$ multiplet in
Table~\ref{tab:MassSpectrumAll}\@.  In particular, using the value of
$M_0^{c\bar c s\bar s}$ from Eqs.~(\ref{eq:ccssBest}) and the
lattice-simulated glue potentials $V(r)$ of
Refs.~\cite{Juge:2002br,Morningstar:2019,Capitani:2018rox}, we
compute
\begin{eqnarray} \label{eq:csDiquark}
m_{\de(cs)} & = & 2080.2 \pm 1.5 \ {\rm MeV} \ {\rm (JKM)} \, ,
\nonumber \\
& = & 2058.5 \pm 1.5 \ {\rm MeV} \ {\rm (CPRRW)} \, .
\end{eqnarray}
The $M_0^{c\bar c s\bar s}$ and $\kappa_{sc}$ values obtained in
Eqs.~(\ref{eq:ccssBest}) and (\ref{eq:csDiquark}) differ rather
little from those in Ref.~\cite{Giron:2020qpb}, in part because the
previous work effectively takes $V_8 \! = \! 0$, and also because the
inputs of Eqs.~(\ref{eq:X3915}) and (\ref{eq:FinalX4140}) have
changed little in the interim.

Feeding the value of nonzero $V_8$ back into the
$c\bar c q\bar q^\prime$ expressions given by Eqs.~(\ref{eq:V8inP})
and (\ref{eq:ccqqParams}), one obtains the $V_8 \! > \! 0$ values of
$M_0^{c\bar c q\bar q^\prime}$, $\kappa_{qc}$, $V_0$, and $P$ given
in Table~\ref{tab:ModelParams}.  Using this value of
$M_0^{c\bar cq\bar q^\prime}$ and the lattice-simulated glue
potentials $V(r)$ of Refs.~\cite{Juge:2002br,Morningstar:2019,
Capitani:2018rox}, we compute
\begin{eqnarray} \label{eq:cqDiquark}
m_{\de(cq)} & = & 1938.0 \pm 0.9 \ {\rm MeV} \ {\rm (JKM)} \, ,
\nonumber \\
& = & 1916.2 \pm 0.9 \ {\rm MeV} \ {\rm (CPRRW)} \, .
\end{eqnarray}
The averaged values for Eqs.~(\ref{eq:csDiquark}) and
(\ref{eq:cqDiquark}) appear in Table~\ref{tab:ModelParams}.

\subsection{$c\bar c q\bar s$ Sector}

The results obtained from the $c\bar c q\bar q^\prime$ and
$c\bar c s\bar s$ $\Sigma^+_g(1S)$ multiplets in the previous two
subsections, with parameters collected in
Table~\ref{tab:ModelParams}, are almost completely sufficient to
predict the entire $c\bar c q\bar s$ $\Sigma^+_g(1S)$ spectrum.
However, as noted in Sec.~\ref{subsec:SU3Mixing}, the fact that the
open-strange states are pure SU(3)$_{\rm flavor}$ octet, while
$c\bar c q\bar q^\prime$ and $c\bar c s\bar s$ are assumed to be
ideally mixed octet-singlet combinations, means that the value of
$M_0^{c\bar c q\bar s}$ extracted using only inputs from the other
sectors is likely to be slightly too low to match observed $Z_{cs}$
masses in Eqs.~(\ref{eq:LHCbZcs}).  On the other hand, the fine
structure obtained in this sector using values of $\kappa_{sc}$,
$\kappa_{sq}$, and $V_8$ from Table~\ref{tab:ModelParams} should be
predicted correctly.

Explicitly, using the $m_{\de(cq)}$ from Eqs.~(\ref{eq:cqDiquark}),
$m_{\de(cs)}$ from Eqs.~(\ref{eq:csDiquark}), and the same
lattice-calculated potentials $V(r)$ as used previously, we compute
\begin{eqnarray}
M_0^{c\bar c q\bar s} & = & 4119.7 \pm 1.7 \ {\rm MeV} \ {\rm (JKM)}
\, , \nonumber \\
& = & 4119.7 \pm 1.7 \ {\rm MeV} \  {\rm (CPRRW)} \, ,
\end{eqnarray}
a remarkably stable result across simulations.  Using this value
along with the other parameters in Table~\ref{tab:ModelParams} in the
$1^+$ expressions of Eqs.~(\ref{eq:OpenStrangeExact}), we
compute
\begin{eqnarray} \label{eq:ZcsNoOffset}
m_{Z_{cs}^{(1)}} & = & 3967.7 \pm 3.4 \ {\rm MeV} \, , \nonumber \\
m_{Z_{cs}^{(2)}} & = & 4136.0 \pm 7.1 \ {\rm MeV} \, , \nonumber \\
m_{Z_{cs}^{(3)}} & = & 4190.3 \pm 3.1 \ {\rm MeV} \, ,
\end{eqnarray}
which are lower than the measured values given in
Eqs.~(\ref{eq:LHCbZcs}).  If, however, we add an offset
\begin{equation} \label{eq:Offset}
\Delta M_0^{c \bar c q\bar s} = 35.3 \pm 6.9 \ {\rm MeV} \, ,
\end{equation}
then the predictions of Eqs.~(\ref{eq:ZcsNoOffset}) become
\begin{eqnarray} \label{eq:ZcsOffset}
m_{Z_{cs}^{(1)}} & = & 4003.0 \pm 7.6 \ {\rm MeV} \, , \nonumber \\
m_{Z_{cs}^{(2)}} & = & 4171.3 \pm 10.3 \ {\rm MeV} \, , \nonumber \\
m_{Z_{cs}^{(3)}} & = & 4225.6 \pm 7.5 \ {\rm MeV} \, ,
\end{eqnarray}
and so $m_{Z_{cs}^{(1)}}$ and $m_{Z_{cs}^{(3)}}$ beautifully match
the observed values in Eqs.~(\ref{eq:LHCbZcs}).

Of course, this model predicts also a third open-strange $1^+$ state
$Z_{cs}^{(2)}$, which is not, as yet, reported by LHCb\@.  In this
regard, we note that the reported mass uncertainty and width of
$Z_{cs}(4220)$ in Eqs.~(\ref{eq:LHCbZcs}) are quite large, meaning
that subsequent analysis might resolve the peak as two states
$Z_{cs}^{(2)}$ and $Z_{cs}^{(3)}$, as was found for the pentaquark
candidates $P_c(4440)$ and $P_c(4457)$ in Ref.~\cite{Aaij:2019vzc}.
Small hints of additional structure may be already visible in the
LHCb results above 4100~MeV (Fig.~3 of \cite{Aaij:2019vzc}, right
inset).  The relative closeness of $Z_{cs}^{(2)}$ and $Z_{cs}^{(3)}$
in mass follows directly in this model, as can be seen from
Eqs.~(\ref{eq:ZcsApprox}), since (Table~\ref{tab:ModelParams})
$\kappa_{qc} \! \ll \! \kappa_{sc}$.  The large mass splitting
$m_{Z_{cs}^{(3)}} - m_{Z_{cs}^{(1)}} \! \approx \! 2\kappa_{sc}
\! > \! 200$~MeV is also explained naturally by the model; in this
regard, note that such a large mixing would not have occurred without
the mixing of strange states between $1^{++}$ and $1^{+-}$
multiplets, as discussed in Sec.~\ref{sec:OpenStrange}.

One other criterion may be useful for disentangling the trio of
$1^+$ states $Z_{cs}^{(i)}$: The eigenvectors of
Eqs.~(\ref{eq:OpenStrangeExact}) couple differently to the heavy- and
light-quark-spin eigenstates $X_1$ [Eqs.~(\ref{eq:SwaveQQ})],
$\tilde Z^\prime$, and $\tilde Z$ [defined in
Eqs.~(\ref{eq:HQbasis})].  Explicitly, in the limit $V_8 \! = \! 0$,
the corresponding eigenvectors with respect to this basis are
\begin{equation}
v^{(1),(2),(3)} = \frac 1 2 \! \left( \begin{array}{c} +\sqrt{2} \\
-1 \\ +1 \end{array} \! \right) , \
\frac 1 2 \! \left( \begin{array}{c} -\sqrt{2} \\ -1 \\ +1
\end{array} \! \right) , \
\frac{1}{\sqrt{2}} \! \left( \begin{array}{c} 0 \\ +1 \\ +1
\end{array} \! \right) .
\end{equation}
Since only $\tilde Z$ has $s_{Q\bar Q} \! = \! 0$, the third
component of each eigenvector indicates the relative strength of the
coupling to $h_c$ ({\it vs.} $J/\psi$) in decays that conserve
heavy-quark spin.  We note that $Z_{cs}^{(2)}$ has the largest such
coupling: 50\% of its decays should be to $s_{Q\bar Q} \! = \! 0$
states.  Therefore, a prediction of this model is not only that
$Z_{cs}(4220)$ resolves into two peaks, but also that the lower state
couples particularly strongly to $h_c$.

Using the values of $M_0^{c\bar c q\bar s}$, $\kappa_{sc}$,
$\kappa_{qc}$, and $V_8$ from Table~\ref{tab:ModelParams} and the
shifted multiplet average mass [using Eq.~(\ref{eq:Offset})],
\begin{equation} \label{eq:M0wOffset}
\tilde{M}_0^{c\bar c q\bar s} \equiv M_0^{c\bar c q\bar s} +
\Delta M_0^{c\bar c q\bar s} = 4155.0 \pm 7.5 \ {\rm MeV} \, ,
\end{equation}
we tabulate mass values for all members of the $\Sigma^+_g(1S)$
$c\bar c q\bar s$ multiplet in Table~\ref{tab:MassSpectrumAll}\@.
\begin{table*}
\caption{Predictions of hidden-charm plus light-quark tetraquark
meson masses (in MeV) for all states in the lowest multiplet
[$\Sigma^+_g(1S)$] of the dynamical diquark model, using the
Hamiltonian parameters of Table~\ref{tab:ModelParams} [and
Eq.~(\ref{eq:M0wOffset}) for $c\bar c q\bar s$].  Observed masses
(used to obtain the Hamiltonian parameters) are exhibited in
boldface.  Uncertainties are obtained by including those on all fit
parameters.}
\centering
\setlength{\extrarowheight}{1.2ex}
\begin{tabular}{ c | c  c | c  c|  c  c | c |  c  c}
\hline\hline
& \multicolumn{4}{c|}{$c\bar{c}q\bar{q^\prime}$} &
\multicolumn{2}{c|}{$c\bar{c}s\bar{s}$} &
\multicolumn{3}{c}{$c\bar{c}q\bar{s}$}\\
\hline
$J^{PC}$ & \multicolumn{2}{c|}{$I=0$} & \multicolumn{2}{c|}{$I=1$}& &
&$J^P$ & & \\
\hline
$0^{++}$ & $3841.9 \pm 13.0$ & $4262.1\pm 17.1$ & $3872.1\pm 9.6$ &
$3988.7 \pm 2.3$ & $\mathbf{3921.7} \pm 4.3$ &
$\mathbf{4350.9}\pm 4.7$ & $0^+$ & $3951.8\pm 7.7$ & $4228.0\pm 10.4$
\\
$1^{+-}$ & $3896.0\pm 5.9$ & $4259.2\pm 17.1$ &
$\mathbf{3887.2}\pm 6.1$ & $\mathbf{4024.4}\pm 3.0$ & $4135.8\pm 3.7$
& $4356.4\pm 3.4$ & $1^+$ & $\mathbf{4003.0}\pm 7.6$ &
$4171.3\pm 10.3$\\
$1^{++}$ & $\mathbf{3872.2}\pm 7.8$ & $\cdots$ & $3993.8\pm 5.6$ &
$\cdots$ & $\mathbf{4146.7}\pm 3.5$ & $\cdots$ & &
$\mathbf{4225.6}\pm 7.5$ & $\cdots$\\
$2^{++}$ & $3923.4\pm 7.8$ & $\cdots$ & $4045.0\pm 5.6$ & $\cdots$ &
$4366.3\pm 3.5$ & $\cdots$ & $2^+$ & $4220.1\pm 7.5$ & $\cdots$\\
\hline
\end{tabular}
\label{tab:MassSpectrumAll}
\end{table*}
\subsection{The $c\bar c uds$ Pentaquark}

The numerical analysis of Ref.~\cite{Giron:2019bcs} builds upon the
proposal of pentaquarks as diquark-triquark bound
states~\cite{Lebed:2015tna}, where the triquark $\bt$ is formed using
a concatenation of the same color-triplet-binding mechanism as
appears within the diquarks:
\begin{equation}
\bt \equiv \left[ \bar Q (q_1 q_2)_{\bar {\bm 3}} \right]_{\bm 3} \,
. 
\end{equation}
The dynamical diquark model then uses the same lattice-calculated
potentials as before~\cite{Juge:2002br,Morningstar:2019,
Capitani:2018rox} to connect the diquark and triquark quasiparticles.
In the original calculations of Ref.~\cite{Giron:2019bcs}, the
pentaquark candidates $P_c(4312)$, $P_c(4380)$, $P_c(4440)$, and
$P_c(4457)$ observed at LHCb~\cite{Aaij:2015tga,Aaij:2019vzc} are
considered as $\de \! = \! (cu)$, $\bt \! = \! [\bar c (ud)]$ bound
states using a coarse analysis:  LHCb identifies $P_c(4380)$ as
having opposite parity to
$P_c(4440)/P_c(4457)$~\cite{Aaij:2015tga},\footnote{It is actually
the original, unresolved $P_c(4450)$ of Ref.~\cite{Aaij:2015tga} that
has opposite parity to $P_c(4380)$, and an assumption of this work
that the two resolved components $P_c(4440)$ and
$P_c(4457)$~\cite{Aaij:2019vzc} share this same parity eigenvalue.}
and such a small splitting between multiplets of opposite parity in
the model makes sense only if $P_c(4380)$ is a high-lying state in
the $P \! = \! -$ multiplet\footnote{The $g$ quantum number is lost
in the asymmetric diquark-tri\-quark case~\cite{Lebed:2017min}.}
$\Sigma^+(1S)$
$\big[ J^P \! = \! \big(\frac 1 2 , \frac 3 2 \big)^- \big]$, while
the other states belong to the $P \! = \! +$ multiplet $\Sigma^+(1P)$
$\big[ J^P \! = \! \big(\frac 1 2 , \frac 3 2 , \frac 5 2 \big)^+ \big]$.

Using $P_c(4312)$ to fix $M_0$ for the $\Sigma^+(1P)$ multiplet,
where~\cite{Aaij:2019vzc}
\begin{equation}
m_{P_c(4312)} = 4311.9 \pm 6.8 \ {\rm MeV} \, ,
\end{equation}
and using $m_{\de = (cq)}$ obtained from the $c\bar c q\bar q^\prime$
states, Ref.~\cite{Giron:2019bcs} computes a value of
$m_{\bt} \! \simeq \! 1.93$~GeV\@.  Repeating the analysis here using
the new value of $m_{\de = (cq)}$ from Eq.~(\ref{eq:cqDiquark}), we
find
\begin{eqnarray}
m_{\bt = \bar c (q_1 q_2)} & = & 1884.6 \pm 7.5 \ {\rm MeV} \
{\rm (JKM)} \, , \nonumber \\
& = & 1866.5 \pm 7.5 \ {\rm MeV} \ {\rm (CPRRW)} \, .
\end{eqnarray}
As noted in Ref.~\cite{Giron:2019bcs}, $m_{\de = (cq)}$ and
$m_{\bt = [\bar c (q_1 q_2)]}$ are quite close in mass; indeed,
$m_{\bt}$ is actually slightly smaller than $m_{\de}$ in the new
calculation.  This peculiarity arises from assigning the lowest
observed $c\bar c q\bar q^\prime$ states to the ground-state
$\Sigma^+_g(1S)$ multiplet but assigning the lowest observed
pentaquark $P_c(4312)$ to the excited $\Sigma^+(1P)$ multiplet.
Should the opposite-parity $P_c(4380)$ disappear from future data,
then $P_c(4312)$ and the other states would become suitable to belong
to $\Sigma^+(1S)$, and $m_{\bt}$ (absorbing what was orbital
excitation energy) would become numerically larger.  Even in the
current circumstance, however, no obvious physical requirement
demands that $m_{\bt} \! > \! m_{\de}$.  Indeed, one may argue that
$\bt$ contains two significant sources of binding energy: within the
diquark $(q_1 q_2)$, and between this diquark and $\bar c$, while
$\de$ possesses only the first type of binding, thereby allowing
$m_{\bt} \! \alt \! m_{\de}$.

Now suppose that the state $P_{cs}(4459)$ of Eqs.~(\ref{eq:Pcs}) is
the open-strange $\Sigma^+(1P)$ analogue to $P_c(4312)$, {\it i.e.},
a $\de$-$\bt$ state with $\de \! = \! (cs)$, where $m_{\de = (cs)}$
is given in Eqs.~(\ref{eq:csDiquark}).  Then we compute
\begin{eqnarray} \label{eq:PcsMass}
M_0^{c\bar c s q_1 q_2} & = & 4441.8 \pm 7.0 \ {\rm MeV} \
{\rm (JKM)} \, , \nonumber \\
& = & 4441.7 \pm 7.0 \ {\rm MeV} \ {\rm (CPRRW)} \, . 
\end{eqnarray}
This is a stunning result, being less than 2$\, \sigma$ lower than
the value in Eqs.~(\ref{eq:Pcs}).  Indeed, no reason apart from
convenience leads one to take the $P_c(4312)$ [as opposed to, say,
$P_c(4457)$] and $P_{cs}(4459)$ masses equal to the $M_0$ values for
their respective $\Sigma^+(1P)$ multiplets, except that they are the
lightest ones known.  A complete analysis would incorporate fine
structure, as is done for the tetraquark sectors, but this exercise
has not yet been carried out in the pentaquark sectors of this model,
due to a lack of experimental clarity on $J^P$ quantum numbers for at
least some of the observed states.  Nevertheless, the result of
Eqs.~(\ref{eq:PcsMass}) shows that a single model can, in fact,
accommodate exotics in all observed flavor sectors.

The development of the dynamical diquark model to date has focused
primarily on spectroscopy, and to a lesser extent on identifying the
dominant quarkonium decay channels.  Detailed quantitative
calculations of strong decay widths, particularly for
open-heavy-flavor channels, have not yet been attempted, because a
precise description of couplings between diquark and hadron-hadron
configurations has not yet been developed.  Qualitative statements to
explain the relative narrowness of exotic states have appeared since
the initial description of the picture in
Refs.~\cite{Brodsky:2014xia,Lebed:2015tna}; they originate from the
significant spatial separation between the diquark or triquark
quasiparticles, which hinders the rearrangement of their component
quarks into color-singlet hadrons.  Such effects may need to be quite
potent in the pentaquarks, since $P_c(4312)$, $P_c(4440)$,
$P_c(4457)$, $P_{cs}(4459)$ all have surprisingly small widths
($\alt \! 20$~MeV).  In addition, the proximity of these states to
$\overline{D}^{(*)} \Sigma_c^{(*)}$ or $\overline{D}^{(*)} \Xi_c$
thresholds has been noted since the original experimental papers,
and predicted earlier in molecular models~\cite{Wu:2010jy,
Wang:2011rga,Yang:2011wz,Karliner:2015ina}.  The next phase of the
development of the model will address the effect of mixing between
diquark configurations and hadron-hadron thresholds, thus providing
critical insight into both the decay properties of exotics and a
connection to the successes of hadronic-molecule pictures.

\section{Conclusions}
\label{sec:Concl}

This work shows that the newly observed hidden-charm, open-strange
exotic-hadron candidates $Z_{cs}(4000)$, $Z_{cs}(4220)$, and
$P_{cs}(4459)$ fit naturally into the dynamical diquark model.
Notably, the same lattice-simulated potential $V(r)$ between two
heavy, color-triplet sources in the lowest Born-Oppenheimer
configuration $\Sigma^+_{(g)}$ is seen to apply to all cases studied
here.

Among tetraquarks, the same Hamiltonian parameters, with numerical
values obtained from the $c\bar c q\bar q^\prime$ ($q$, $q^\prime$
being $u$ or $d$) and $c\bar c s\bar s$ exotic $\Sigma^+_g(1S)$
multiplets, successfully predict masses in the $c\bar c q\bar s$
sector.  In particular, the large $Z_{cs}(4220)$-$Z_{cs}(4000)$ mass
splitting emerges naturally as consequences of both the large $(cs)$
diquark internal spin-spin coupling $\kappa_{sc}$ and the mixing of
open-strange members of $J^{PC} \! = \! 1^{++}$ and $1^{+-}$
multiplets, the latter an effect seen in conventional hadron physics
for strange mesons such as $K_{1A}$ and $K_{1B}$.  The model also
predicts a third $1^+$ $Z_{cs}$ state lying not far below 4200~MeV\@.

The overall multiplet-average mass $M_0$ for $c\bar c q\bar s$ states
also receives a shift modification compared to those for
$c\bar c q\bar q^\prime$ and $c\bar c s\bar s$ states, since the
former are pure SU(3)$_{\rm flavor}$ octet states, while states in
the latter sets are assumed in the numerical analysis to be ideally
mixed octet-singlet combinations.  The size of this shift
$\Delta M_0^{c\bar c q\bar s}$ is found to be numerically not large,
at most a few 10's of MeV\@.

In addition, the model in all sectors has been expanded to allow not
only a $\pi$-like interaction operator between the diquarks (as in
previous studies), but an $\eta$-like interaction operator as well.
The numerical size of the coefficient $V_8$ of this operator is found
to be much smaller than that ($V_0$) for $\pi$-like interactions, and
has a fairly minimal effect on the hadron spectra.  Mass predictions
for all states in the $\Sigma^+_g(1S)$ multiplet for each flavor
content are presented.

Among pentaquarks, a crude calculation taking the nonstrange
$P_c(4312)$ as a base state for the positive-parity multiplet
$\Sigma^+(1P)$ constructed of a diquark-triquark pair
$(cq)[\bar c (ud)]$, and replacing the $(cq)$ diquark with a $(cs)$
diquark, produces a state very close in mass to that of
$P_{cs}(4459)$.

As new exotic hadrons continue to be uncovered---a rather safe
expectation, considering the rate of observational advances over the
past few years---more opportunities for sharpening our understanding
of their mass spectrum and transitions will emerge.  Whether or not a
diquark-based spectrum provides the eventual global picture for these
states, the dynamical diquark model supplies a definite road map for
the sort of spectrum to expect.
\vspace{1em}

\begin{acknowledgments}
This work was supported by the National Science Foundation (NSF) under 
Grant No.\ PHY-1803912.  We thank T.~Skwarnicki for helpful insight
into the LHCb results.
\end{acknowledgments}

\bibliographystyle{apsrev4-1}
\bibliography{diquark}
\end{document}